\documentclass[twocolumn,pra,aps,raggedbottom, superscriptaddress]{revtex4-2}
\usepackage{times}
\usepackage[table,xcdraw]{xcolor}
\usepackage{amsmath,amsfonts,amssymb,graphicx,hyperref,epstopdf,xcolor,tikz,scalerel,natbib,array}
\usepackage{mathptmx,textcomp,braket,diagbox}
\usepackage[T1]{fontenc}
\usepackage[utf8]{inputenc}
\usepackage{multirow}
\usepackage{array}

\definecolor{lime}{HTML}{A6CE39}
\DeclareRobustCommand{\orcidicon}
{
	\begin{tikzpicture} 
	\draw[lime, fill=lime] (0,0) circle [radius=0.15] node[white] {{\fontfamily{qag}\selectfont \tiny ID}};
	\draw[white, fill=white] (-0.0625,0.095) 	circle [radius=0.007];
	\end{tikzpicture}
	\hspace{-2.2mm}
}
\newcommand\orcidID[1]{\href{https://orcid.org/#1}{\orcidicon}}

\newcommand{\be}{\begin {equation}}
\newcommand{\ee}{\end {equation}}
\newcommand{\beqa}{\begin {eqnarray}}
\newcommand{\eeqa}{\end {eqnarray}}
\newcommand{\mb}{\mathbf}
\newcommand{\Sch}{Schr\"odinger }
\newcommand{\Exp}[1]{\text{e}^{#1}}
\newcommand{\tioni}{t_\text{ioni}}
\newcommand{\treco}{t_\text{reco}}

\hypersetup{colorlinks,citecolor=blue,filecolor=black,linkcolor=blue,urlcolor=blue}

\begin{document}

\title{High-harmonic generation as a tunneling delay probe}
  
\author{Amol R. Holkundkar\orcidID{0000-0003-3889-0910}}
\email[E-mail: ]{amol@holkundkar.in}
\affiliation{Department of Physics, Birla Institute of Technology and Science - Pilani, Rajasthan, 333031, India.}
 
\date{\today}

\begin{abstract}
We investigate the feasibility of using high-harmonic generation (HHG) as a complementary probe of tunneling delay in strong-field ionization. By combining time--frequency analysis of HHG spectra obtained from full time-dependent Schrödinger equation (TDSE) simulations with classical three-step-model (TSM) trajectories, we extract an effective tunneling delay associated with electron motion through the laser-suppressed Coulomb barrier. The analysis is carried out for Hydrogen, Helium, and Argon atoms over a range of laser wavelengths and peak intensities within the tunneling regime. The extracted delay exhibits a systematic dependence on the instantaneous field strength and barrier width at the ionization time, and follows the expected $\tau_d \propto 1/\sqrt{I_0}$ scaling consistent with Keldysh--Rutherford-type models and attoclock observations. When recast in terms of the Keldysh parameter, the tunneling delay collapses onto a near-universal trend across different atomic species. While HHG does not provide a direct measurement of tunneling time, the present results demonstrate that it can serve as a robust, internally consistent diagnostic of tunneling dynamics, offering an independent and complementary perspective to established attoclock techniques.
\end{abstract}

\maketitle

\section{Introduction}

Strong-field ionization driven by intense laser pulses has become a central paradigm in attosecond science, enabling direct access to electron dynamics on sub-femtosecond timescales. One of the most influential conceptual frameworks in this context is the classical three-step model (TSM), which successfully describes high-harmonic generation (HHG) as a sequence of tunneling ionization, laser-driven propagation in the continuum, and recombination with the parent ion. Despite its classical simplicity, the TSM has proven remarkably successful in predicting harmonic cutoffs, emission times, and many qualitative features observed in experiments \cite{Corkum1993,Lewenstein1994,KrauszIvanov2009}.

While the recombination and propagation steps of the TSM are comparatively well understood, the temporal aspects of the tunneling step remain a subject of active debate. In particular, the question of whether tunneling occurs instantaneously or over a finite time interval has motivated extensive experimental and theoretical investigations \cite{Sainadh2019,Sainadh2019_new, Torlina2015,Landsman_14,Pfeiffer2012,Hofmann_2019_JMO,Camus2020,Ni2020PRL,Ni2021NatCommun}. Attoclock experiments, based on angular streaking in elliptically polarized laser fields, have emerged as one of the most direct approaches to probing tunneling delay. Early measurements demonstrated non-zero angular offsets that were interpreted as signatures of finite tunneling times \cite{Pfeiffer2012}, sparking a sustained effort to clarify their physical origin.

Subsequent experimental advances have significantly refined attoclock measurements, culminating in high-precision studies on atomic hydrogen, which minimizes multielectron and structural complications \cite{Sainadh2019,Sainadh2019_new}. These experiments, together with measurements on more complex targets, have firmly established that the observed offsets depend sensitively on laser intensity, wavelength, and the Coulomb interaction with the parent ion \cite{PhysRevLett.117.023002,Camus2020,Ni2020PRL}. At the same time, they have highlighted the subtle interplay between tunneling dynamics, exit-channel effects, and post-ionization Coulomb focusing.

On the theoretical side, substantial progress has been made in interpreting attoclock measurements. Quantum trajectory and Coulomb-corrected models have shown that the measured angular offsets cannot be attributed solely to a single “tunneling time,” but instead reflect a combination of under-the-barrier dynamics and continuum propagation effects \cite{Torlina2015,RostSaalmann2019,Kaushal2016_PRA,PhysRevA.90.012116}. In this context, the Keldysh–Rutherford (KR) model has provided a particularly transparent description, predicting a tunneling delay that scales approximately as $\tau \propto 1/\sqrt{I_0}$ in the tunneling regime \cite{Bray2018}. More recent studies have further emphasized the role of nonadiabatic corrections and wavelength-dependent effects, especially at long wavelengths and low Keldysh parameters \cite{Pisanty2020,Eckart2021,Liu2021,Zhang2022}.

In parallel with these developments, HHG has evolved into a powerful, self-referencing probe of strong-field electron dynamics. Time–frequency analysis of HHG spectra enables the association of emitted harmonics with specific ionization and recombination times, providing a direct link between full quantum-mechanical simulations and classical electron trajectories \cite{Smirnova2009,Chirala2010}. Unlike attoclock techniques, HHG experiments are typically performed with linearly polarized fields and are inherently sensitive to recombination dynamics. As a result, HHG does not provide a direct chronoscopic measurement of tunneling time. Nevertheless, the close correspondence between HHG emission times obtained from time-dependent Schrödinger equation (TDSE) simulations and classical TSM predictions suggests that HHG may encode indirect but robust information about the tunneling step itself.

Motivated by this perspective, the present work explores HHG as a supplementary diagnostic of tunneling delay rather than as an alternative to attoclock measurements. By combining TDSE simulations with time–frequency (Gabor) analysis and classical trajectory calculations, we extract an effective tunneling delay associated with electron motion through the laser-suppressed Coulomb barrier. The analysis is carried out for Hydrogen, Helium, and Argon atoms over a range of laser intensities and wavelengths within the tunneling regime. Particular emphasis is placed on identifying systematic trends with peak intensity, instantaneous field strength at ionization, and the Keldysh parameter.

Before proceeding, it is important to clarify the scope and interpretation of the tunneling-delay diagnostic discussed in this work. The quantity $\tau_d$ extracted here should not be regarded as a tunneling-time observable in a strict quantum-mechanical sense, nor as a direct measurement analogous to attoclock experiments. Instead, $\tau_d$ represents an effective, model-dependent delay inferred within a combined TDSE and classical-trajectory framework, where the tunneling step is characterized through the instantaneous barrier geometry and the timing information encoded in HHG emission. The present approach is therefore intended as a diagnostic and comparative tool, aimed at identifying systematic trends and internal consistency rather than providing a unique definition of tunneling time. Agreement with attoclock-based scaling laws and Keldysh-parameter dependence should be interpreted as a consistency check across complementary frameworks, rather than as a validation or replacement of existing tunneling-time measurements.

The manuscript is organized as follows. In Sec. \ref{sec2}, we outline the theoretical framework and numerical methodology used to solve the time-dependent Schrödinger equation and to extract timing information from high-harmonic generation. Section \ref{sec3} presents the results and discussion, where the tunneling-delay diagnostic is analyzed as a function of laser intensity, wavelength, atomic species, and barrier geometry. In Sec. \ref{sec4}, the main findings are summarized and recast in terms of the Keldysh parameter to highlight universal trends across different systems. Finally, Sec. \ref{sec5} contains the concluding remarks and perspectives.

\section{Theoretical and Numerical Considerations}

\label{sec2} 

The time-dependent \Sch equation (TDSE) under single-active-electron (SAE) approximation is numerically solved using the time-dependent generalized pseudospectral method (TDGPS) method \cite{TONG1997119}. The interaction of the 4 cycles, linearly polarized laser having $\sin^2$ envelope is studied with Hydrogen, Helium and Argon atomic targets, and the tunneling delays is estimated by varying the wavelength and the peak intensity of the laser pulse. The atomic units are used throughout the manuscript, i.e. $|e| = m_e = \hbar = 1$. The TDSE in the length gauge is written as:
\be
i \frac{\partial}{\partial t} \ket{\psi(\mb{r},t)} = [H_\text{o} + H_\text{L}(t)] \ket{\psi(\mb{r},t)},
\label{tdse0}
\ee
  
where, $H_\text{o} = -\nabla^2/2 + V(r)$ is the field free Hamiltonian and $H_\text{L}(t) = \mb{r}\cdot\mb{E}(t)$ is the interaction Hamiltonian in the length gauge, with $\mb{E}(t)$ being the temporal profile of driving laser electric field under dipole approximation and is given as,
\be
\mb{E}(t) = E_0 \sin(\omega_0 t+\phi) \sin^2\Big(\frac{\pi t}{T}\Big) \hat{z}.
\label{laser}
\ee
Here, $E_0 \text{[a.u.]} \sim 5.342\times 10^{-9} \sqrt{I_0}$ being the field amplitude, with peak intensity $I_0$ in W/cm$^2$, $\omega_0\text{[a.u.]} \sim 45.5633/\lambda$ is the frequency of the laser with $\lambda$ being the wavelength in nanometers. The CEP of the pulse $\phi = 0$ is considered, and $T = 4 \tau_0$ is the duration of the laser pulse unless otherwise mentioned, with $\tau_0 \text{[a.u.]} = 2\pi/\omega_0$ being the duration of the one cycle corresponding to respective wavelength. 

The atomic Coulomb potential $V(r)$ in the $H_\text{o}$ is modeled under single active electron (SAE) approximation by an empirical expression given by \cite{Tong_2005}:
\be V(r) = -\frac{1 + a_1\ \Exp{-a_2 r} + a_3\ r\ \Exp{-a_4 r} + a_5\ \Exp{-a_6 r}}{r} \label{potential}\ee
The values of the coefficients $a_i$'s for atomic species Hydrogen, Helium, and Argon are tabulated in Ref. \cite{Tong_2005}. This empirical expression of $V(r)$ is based on the self-interaction free density functional theory under SAE approximation. The advantage of using the TDGPS method with this atomic potential is that, we do not need to use any soft-core kind of potential when solving the TDSE in the Cartesian coordinates. In TDGPS, the radial domain $r = [0,R_{max}]$ is mapped on the range $s =  [-1,1]$ which is further discretized using the \textit{roots of the Legendre polynomial}. As a result no matter how large the radial points we consider between $s \equiv [-1,1]$ the radial point $r = 0$ corresponding to $s = -1$ is never incorporated in the simulation domain, and hence no soft-core type of approximation is needed. This accurate consideration of the model potential enables us to calculate the Ionization potential and other host of quantities with great precision. For example, in this work we adopted the radial simulation domain of $R_{\text{max}} \sim 250$ atomic units (a.u.), with the last $30$ a.u. utilized as a masking region to absorb the outgoing wavefunction \cite{Holkundkar2023_PhysLettA, Rajpoot2023_JPhysB}. For this we employed $N = 1200$ non-uniform grid points (roots of Legendre polynomials) along the radial direction, $L = 150$ as the maximum angular momentum, and a simulation time step of $0.05$ a.u is considered. The convergence was tested concerning the spatial grid and time step.  

In order to test the validity of our TDSE solver, we have tabulated and compared the ionization potential of Hydrogen, Helium, and Argon in Table - \ref{table1}, along with the expectation values of $\braket{r^k}$ with $k=-2,-1,1$ for the Hydrogen atom (as it can be directly compared with the theoretical values) using the eigen-functions of the field free Hamiltonian. The numerically obtained values are found to be in excellent agreement with the theoretical predictions \cite{BransdesnBook}, furthermore it also suggests the appropriate numerical evaluation of the initial ($t = 0$) electronic wavefunction $\ket{\psi_{n\ell}}$ for Eq. \eqref{tdse0}, which are nothing but the eigen-functions of the field-free Hamiltonian. In this work, we are using the linearly polarized laser pulse, and hence the temporal evolution of the wavefunction would be symmetric along the azimuthal direction ($m=0$ case) for all the results presented.  

\bgroup
\def\arraystretch{1.3}%
\begin{table}[!tbp]
\caption{The numerically calculated (through our simulations) ionization potential ($I_p$) for the Hydrogen, Helium and Argon  are compared with theoretical and the experimental results. The Hydrogen atom problem can be solved analytically \cite{BransdesnBook}, and hence we presented the numerically calculated values of the expectation values of field free Hamiltonian [$E_n = \braket{\psi_{n\ell}|H_\text{o}|\psi_{n\ell}}$] for different quantum numbers $[n,\ell]$. Furthermore, $\braket{\psi_{n\ell}|r^k|\psi_{n\ell}}$ with $k=-2,-1,1$ are also shown. These numerically calculated results are in good agreement with the theoretical predictions \cite{BransdesnBook}.} 
\vspace{0.2cm}
\begin{tabular}{c c c c c }
 \cline{2-5}
 \cline{2-5}
   & $I_p$ [a.u.] & Hydrogen & Helium & Argon \\
 \cline{2-5}
   & Numerical & $0.5$ & $0.9038$ & $0.5797$ \\
   & Theoretical \cite{BransdesnBook} & $0.5$ & $0.9037$ & $-$ \\
   & NIST\footnote{\url{https://physics.nist.gov/}} \cite{NIST_ASD} & $0.4997$ & $0.9036$ & $	0.5792$ \\
   \cline{2-5}
  \hline
  \hline
   
 Hydrogen\\$\ket{\psi_{n\ell}}$  & $E_n$ [a.u.] &$\braket{\psi_{n\ell}|r|\psi_{n\ell}}$ & $\braket{\psi_{n\ell}|\frac{1}{r}|\psi_{n\ell}}$ & $\braket{\psi_{n\ell}|\frac{1}{r^2}|\psi_{n\ell}}$ \\ \hline
 $\ket{\psi_{1s}}$ & $-0.5$ & $1.5$ & $1$ & 1.99943\\
 $\ket{\psi_{2s}}$ & $-0.125$ & $5.\overline{999}$ & $0.25$ & 0.24993\\
 $\ket{\psi_{2p}}$ & $-0.125$ & $5$ & $0.25$ & $0.08\overline{333}$ \\
 \hline
 \hline   
\end{tabular}

\label{table1}
\end{table}
\egroup
  
After solving the Time-Dependent Schrödinger Equation (TDSE) using the TDGPS method, the time-dependent dipole acceleration $\mb{a}(t)$ is evaluated according to the Ehrenfest theorem \cite{sandPRL_1999}:
\be
\mb{a}(t) = - \Big\langle \psi(\mb{r},t) \Big| \frac{\partial V(r)}{\partial r} + \mb{E}(t) \Big| \psi(\mb{r},t) \Big\rangle.
\label{dip_acc}
\ee
Subsequently, the harmonic spectra are obtained by Fourier transforming $\mb{a}(t)$:
\be S(\Omega) \sim |a(\Omega)|^2, \label{hhg_spectra}\ee 
where $a(\Omega)$ represents the Fourier transform of the dipole acceleration. The time-frequency analysis of the dipole acceleration [refer Eq. \eqref{dip_acc}] is carried out by the state-of-the-art numerical library `fCWT' \cite{fcwt}.

\section{Results and Discussions}
\label{sec3}

In this work we have studied the interaction of the linearly polarized laser with Hydrogen, Helium and Argon atomic species. Within the present classical-trajectory framework, the time required to traverse the barrier width $\ell_b$ is identified as an effective tunneling-delay parameter $\tau_d$. All subsequent references to tunneling delay in this manuscript should be understood in this diagnostic sense.

\subsection{Methodology}
\label{sec3a}

\begin{figure}[t!]
\includegraphics[width=\columnwidth]{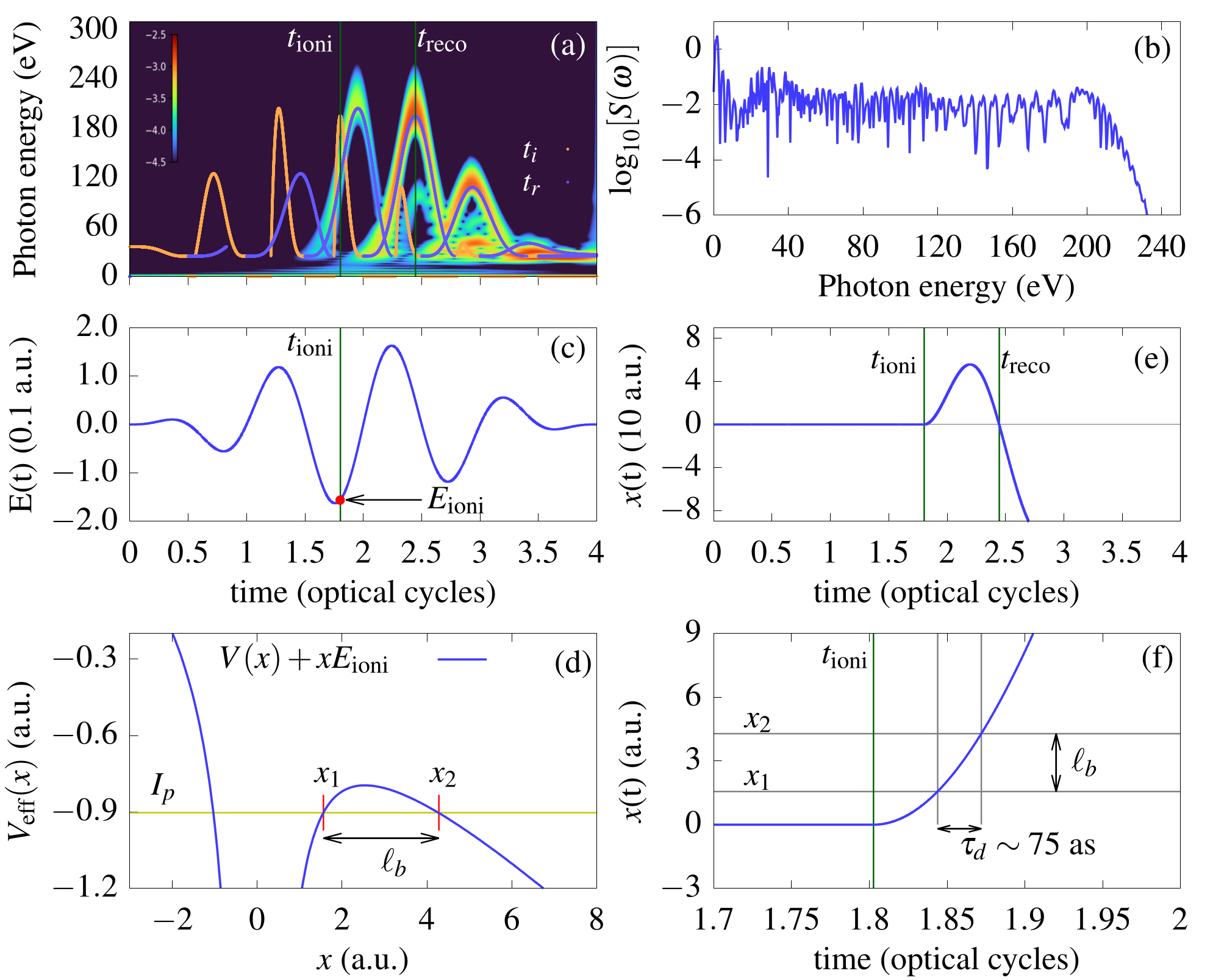}
 \caption{The time-frequency analysis of HHG for the case of Helium atom is calculated along with the ionization and the recombination times as estimated using classical three-step-model (a), along with the HHG spectra in (b). Temporal profile of the laser pulse is shown in (c), such that $E(\tioni) \equiv E_{\rm ioni}$ with $\tioni \sim 1.8 \tau_0$ being the ionization time associated with the emission of the highest energy photon at recombination time $\treco \sim 2.45 \tau_0$. The 1D effective potential $V_{\rm eff}(x) = V(x) + x E_{\rm ioni}$  is presented in (d), and the intersections with the ionization potential ($I_p \sim -0.9038$ a.u.) are marked as $x_1$ and $x_2$. The classical trajectory of the electron is shown in (e) after the ionization at $\tioni$. The zoomed version of the (e) is presented in (f). The ionization time $\tioni$ is shown by the vertical line in (a), (c), (e) and (f). The time ($\tau_d$) electron takes to cover $\ell_b$ is estimated from (f). Laser with pulse duration 4 cycle (sin$^2$ profile), 800 nm with peak intensity of $10^{15}$ W/cm$^2$ is used here. }
\label{fig1}
\end{figure}

In order to elucidate on the methodology we employed to probe the tunneling delay, let us first explore the interaction of laser [$\lambda = 800$ nm, $I_0 = 10^{15}$ W/cm$^2$] with the ground state of the Helium ($I_p \sim 0.9038$ a.u.) atom. The time frequency analysis of the HHG for the same is presented in Fig. \ref{fig1}(a), along with the classical ionization and recombination energies. These classically estimated energies is up-shifted by the ionization energy of the He atom. The typical HHG spectra [refer Eq. \eqref{hhg_spectra}] is shown in Fig. \ref{fig1}(b). The ionization (recombination) time corresponding to the maximum harmonic energy emission is denoted by $\tioni \sim 1.8 \tau_0$ ($\treco \sim 2.45 \tau_0$) and respectively shown by vertical lines in Fig. \ref{fig1}(a). The laser field profile is presented in Fig. \ref{fig1}(c). The field strength at $\tioni$ i.e. $E(\tioni) \equiv E_\text{ioni}$ is obtained from Fig. \ref{fig1}(c). The classical trajectories calculation as shown in Fig. \ref{fig1}(a) are carried out in 1D only, however the TDSE is solved in 2D with azimuthal symmetry with spherically symmetric potential. As a result, the instantaneous effective potential is considered to be $V_\text{eff}(x) = V(x) + x E_\text{ioni}$, where field free potential $V(x)$ is given by Eq. \eqref{potential} in shown in Fig. \ref{fig1}(d). 

A horizontal line corresponding the ionization energy is also shown in Fig. \ref{fig1}(d), which intersects the $V_\text{eff}(x)$ at two points namely $x_1$ and $x_2$. The distance $\ell_b \equiv x_2 -x_1$ is designated as the `Barrier Width'. The classical trajectory $x(t)$ is calculated [Fig. \ref{fig1}(e)] by solving the differential equation $\ddot{x} = -E(t)$ with the initial condition $x(0) = \dot{x}(0) = 0$. In Fig. \ref{fig1}(e), it can be seen that  $x(t < \tioni) = 0$, meaning the electron is not ionized till $\tioni$, however it recombines at $\treco$ which corresponds to the emission of the highest and brightest burst of photon energy [refer Fig. \ref{fig1}(a)]. Furthermore, the zoomed version of the trajectory around $\tioni$ is presented in Fig. \ref{fig1}(f). It is then inferred from the Fig. \ref{fig1}(f), that the time required to cover the distance $\ell_b$ would be refereed as the `\textit{tunneling delay}' $\tau_d$, which in this case is found to be $\tau_d \sim 75$ as. 

\begin{figure}[t]
   \includegraphics[width=\columnwidth]{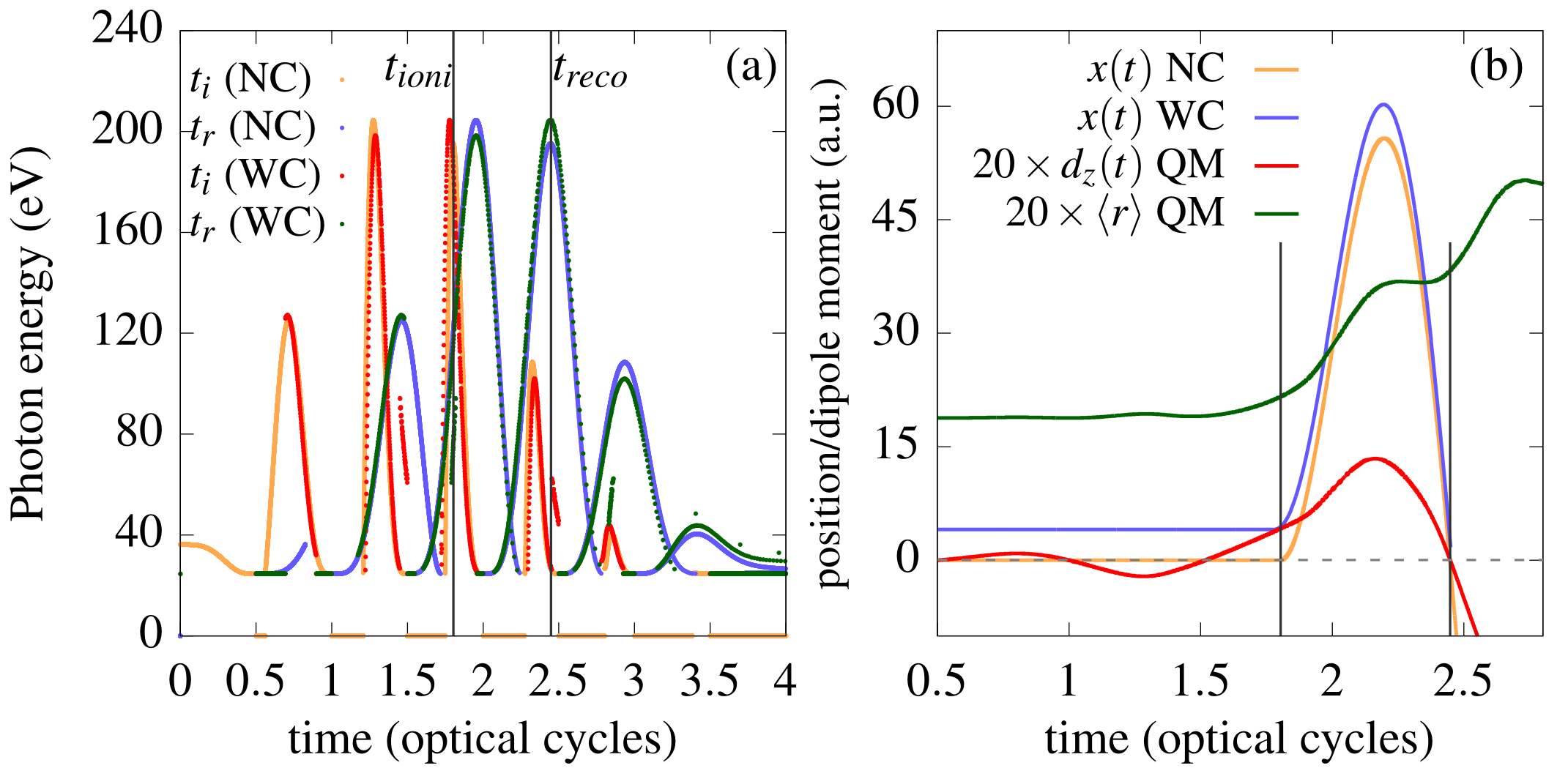}
   \caption{In (a), the ionization and the recombination times estimated using classical three-step-model are compared for the cases when the Couloumb potential is not included (NC) and when it is included (WC) while solving the classical equation of motion. All the physical parameters are same as considered in Fig. \ref{fig1}. The corresponding classical trajectory associated with the highest photon energy emission is also compared for these NC and WC cases (b). In (b), we have also presented the scaled up temporal profile of the dipole moment along the laser polarization direction, $d_z$, and also the expectation value of the radial coordinate i.e. $\braket{r}$. The $d_z$ and $\braket{r}$ is calculated though our TDSE solver. The $t_\text{ioni} \sim 1.8 \tau_0$ and $t_\text{reco} \sim 2.45 \tau_0$ are also shown by two vertical lines in both (a) and (b).  }
   \label{fig2} 
 \end{figure}

As we know that the classical framework is not appropriate to study the ionization dynamics as the electron (SAE) is bound to the nucleous, however the success of the classical TSM to predict the harmonic cutoffs is unpredented. The ionization and recombination times for the classical TSM are obtained by solving the Newton's equation of motion in atomic units i.e. $\ddot{x} = - E(t)$. The initial condition to solve this equation is considered to be $x(t = 0) = \dot{x}(t = 0) = 0$. Here, the Couloumb potential is not included while solving the dynamics. However, in order to explore the effect of the Couloumb potential on the dynamics of the electron in the continum [for $x > x_2$ in Fig. \ref{fig1}(d)], we compare the results of the classical three-step model (TSM) for two cases: one in which the Coulomb potential is neglected (NC) and another in which it is explicitly included (WC). The corresponding results are shown in Fig.~\ref{fig2}(a). For the WC case, the electron is initialized at $x = x_2$, i.e., just outside the potential barrier, and the return time and energy are determined when the electron revisits the same position, since the region $x < x_2$ is classically forbidden. As the ionization time varies, the value of $x_2$ changes accordingly, reflecting the time dependence of the laser-suppressed potential barrier.

As seen in Fig.~\ref{fig2}(a), the ionization time in the WC case is slightly reduced compared to the NC case, owing to the attractive nature of the Coulomb potential, which modifies the early-stage electron motion. Notably, the time at which the maximum return energy is reached-consistent with the TDSE result shown in Fig.~\ref{fig1}(a)—is identical for both NC and WC cases. The maximum return energy in the WC case is marginally higher than in the NC case due to the additional Coulomb acceleration as the electron approaches the parent ion. Apart from these small differences, the overall return-energy predictions of the NC and WC models are in close agreement.

We have also compared the classical trajectories for the NC and WC cases with the temporal evolution of the dipole moment along the laser-polarization direction, $d_z(t)$, and the expectation value of the position, $\langle r(t) \rangle$, as obtained from the TDSE calculations [see Fig.~\ref{fig2}(b)]. It can be seen that the recombination time, $t_{\mathrm{reco}} \sim 2.45\,\tau_0$, is corroborated by the temporal behavior of both $d_z(t)$ and $\langle r(t) \rangle$, which exhibit extrema at the same instant. This agreement provides an internal consistency check between the classical trajectory analysis and the underlying quantum dynamics. 

The consistency between the ionization and recombination times inferred from the TDSE and those obtained from the classical three-step model indicates that the latter can be employed as an effective framework in which the Coulomb potential acts as a background influence on the continuum dynamics. As shown in the following sections, tunneling-delay estimates extracted within this framework are found to be consistent with established attoclock-based experimental and theoretical trends across different atomic species and laser parameters.

Taken together, the results presented in Figs. \ref{fig1} and \ref{fig2} demonstrate that the key quantities governing the present analysis are the ionization time $t_{\mathrm{ioni}}$ and the recombination time $t_{\mathrm{reco}}$. These times are found to be in excellent agreement when extracted independently from the full TDSE-based time--frequency analysis of the HHG signal and from classical three-step-model (TSM) trajectories. The comparison between classical trajectories with and without the Coulomb potential further shows that, while the Coulomb interaction slightly modifies the return energy, it does not alter the timing of ionization or recombination. This robustness of the timing against Coulomb effects indicates that the predictive capability of the classical framework is not compromised for the purpose of extracting an effective tunneling-delay diagnostic. Consequently, once $t_{\mathrm{ioni}}$ and $t_{\mathrm{reco}}$ are fixed by the HHG emission dynamics, the tunneling delay can be consistently inferred within the classical picture, with the Coulomb potential acting primarily as a background influence rather than a source of timing modification.

It should be also noted that while employing the classical framework to estimate the tunnelling delay, the adiabatic approximation is made, i.e. it is assumed that the instantaneous electric field of the laser pulse is not changing during the passage of the electron through the potential barrier of width $\ell_b$, and hence the $V_\text{eff}(x)$ is not altered during this time interval. The validity of this approximation is checked by varying the wavelength of the laser pulse, which is discussed next. 
 
 The extraction of the tunneling-delay diagnostic relies on a quasi-static (adiabatic) approximation, wherein the effective barrier width is evaluated using the instantaneous electric field at the ionization time. While the laser field does, in principle, evolve during the electron’s under-the-barrier motion, this approximation is justified in the present parameter regime. The extracted tunneling delays are on the order of a few tens to a few hundred attoseconds, which are much shorter than the optical cycle of the driving field for the wavelengths considered. Consequently, the change in the electric field during the barrier transit is small, and the effective potential can be treated as approximately static. This assumption is further supported by the observed agreement between TDSE-based and classical ionization and recombination times, as well as by the improved consistency of the results at longer wavelengths and higher intensities, where nonadiabatic effects are expected to be reduced.

 \subsection{Effect of peak intensity and wavelength on $\tau_d$}
 
 The effect of the peak intensity and the wavelength on the tunneling delay is explored in the following. The time-frequency analysis and the classical ionization and recombination time for the Helium for the wavelengths 800 (1200) nm and peak intensities $1.5 (0.5) \times 10^{15}$ W/cm$^2$ are presented respectively in Fig. \ref{fig2}(a) and (b), and the variation of the tunneling delay with the peak intensity for different wavelengths are shown in Fig. \ref{fig2}(c). As we are using the same field envelope [refer Eq. \eqref{laser}] for all the results presented in this work, it is understandable  that the  ionization (recombination) time corresponding to the highest and the brightest harmonic energy is observed to be at $\tioni \sim 1.8\tau_0$ ($\treco \sim 2.45\tau_0$) with $\tau_0$ being the duration of one cycle corresponding to respective wavelengths. As can be seen from Fig. \ref{fig1}(a) and Fig. \ref{fig2}(a-b), that these $\tioni$ and $\treco$ obtained by solving the 2D TDSE are found to be in excellent agreement with the simple 1D classical trajectory based calculations even for different laser wavelengths and the peak intensities.

\begin{figure}[b]
\includegraphics[width=\columnwidth]{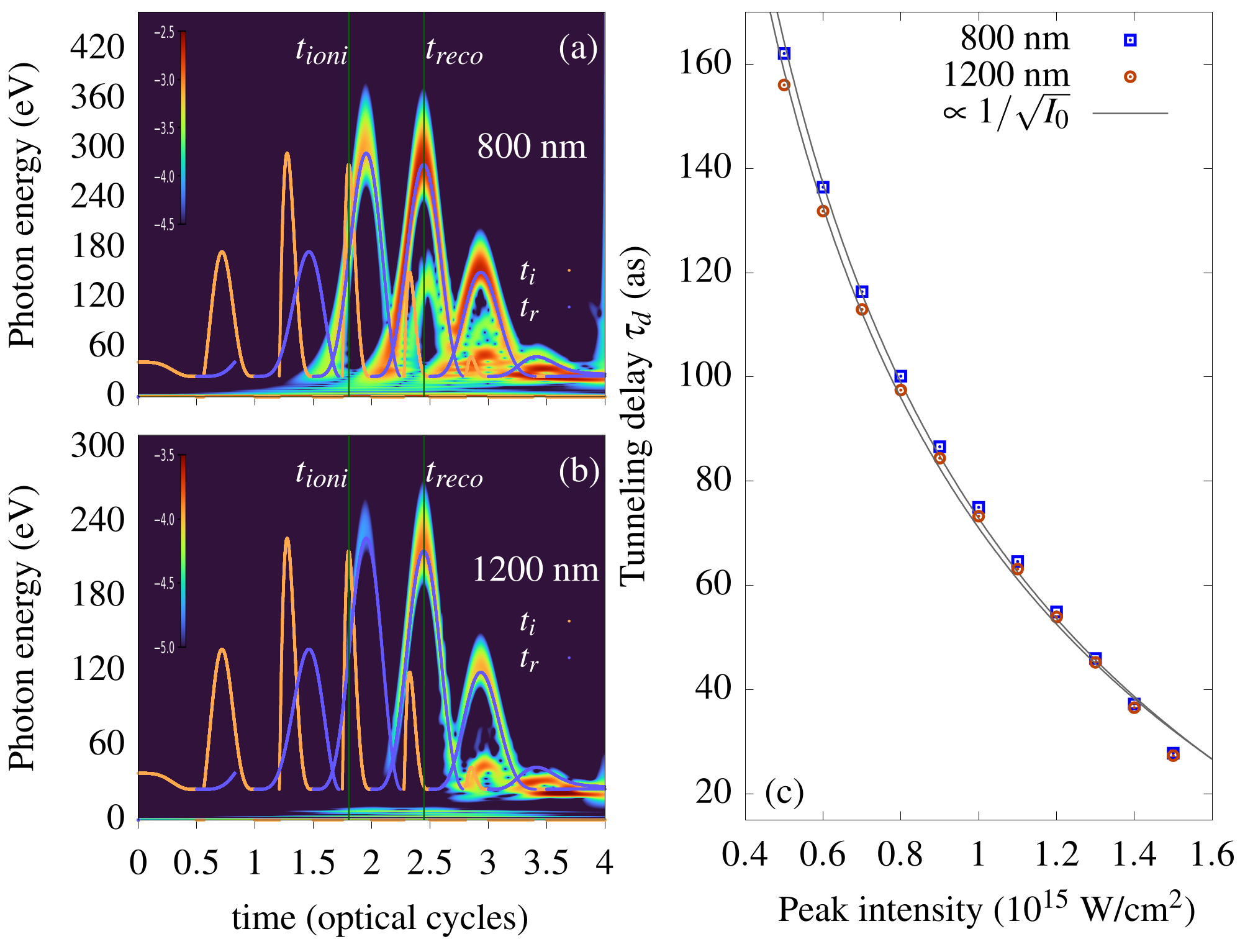}
\caption{The time-frequency and the classical trajectory analysis of the HHG for Helium is presented for wavelengths (intensity) as 800 nm ($1.5\times 10^{15}$ W/cm$^2$) in (a) and 1200 nm ($0.5\times 10^{15}$ W/cm$^2$) in (b). The temporal axis is plotted in terms of their respective optical cycles corresponding to 800 nm (a) and 1200 nm (b). The tunneling delay as a function of the peak intensity of the laser pulse is compared for 800 nm and 1200 nm wavelength cases. The laser pulse duration is considerd to be 4 cycles of respective wavelengths. The solid curve in (c) shows the $\propto 1/\sqrt{I_0}$ scaling for both the wavelength cases. }
\label{fig3}
\end{figure}
 
The methodology of calculation of the tunneling delay is already discussed in Sec. \ref{sec3a}, using the same, the variation of the tunneling delay with the peak laser intensity are calculated for different wavelengths are presented in  Fig. \ref{fig3}(c). It can be observed from the Fig. \ref{fig3}(c), that with peak laser intensity the tunneling delay reduces, which can be be understood from the fact that, with increase in the $I_0$, the corresponding $E_\text{ioni}$ will also increase and from the knowledge of $V_\text{eff}(x)$ it can be seen that the $\ell_b$ would reduce accordingly. Furthermore, it is observed that for each wavelength case the tunneling delay is observed to follow $\tau_d \propto 1/\sqrt{I_0}$ scaling which is consistent with the results obtained using the Keldysh-Rutherford (KR) model where the tunneling delay or the `offset' angles in the `attoclock' terminology is estimated to be proportional to the $1/\sqrt{I_0}$ \cite{Bray2018}. Though the KR model is proposed for the Hydrogen atom, but it can be extended to the cases where the SAE approximation is made with the spherically symmetric potential [refer Eq. \eqref{potential}] like the Hydrogen atom. 

Furthermore, it is also observed that the difference between the tunneling delay $\tau_d$ for 800 nm and 1200 nm lasers decreases gradually from $\sim 6$ as for the lowest intensity value to almost zero as the peak intensity increases. This can be understood from the adiabatic approximation used in the analysis. For lower intensities the barrier width would be more and for shorter wavelengths the width become even longer because during the passage of the electron the field strength further reduces in the same cycle resulting in further increase in the barrier width. For longer wavelengths, the field strength does not change drastically during the passage of the electron through the barrier and hence the barrier width remains almost constant. However, as the intensity increases the barrier width would be smaller and the exit velocity would be higher, as a result the minor change in the barrier width caused by different wavelengths becomes redundant.

 \begin{figure}[t]
   \includegraphics[width=\columnwidth]{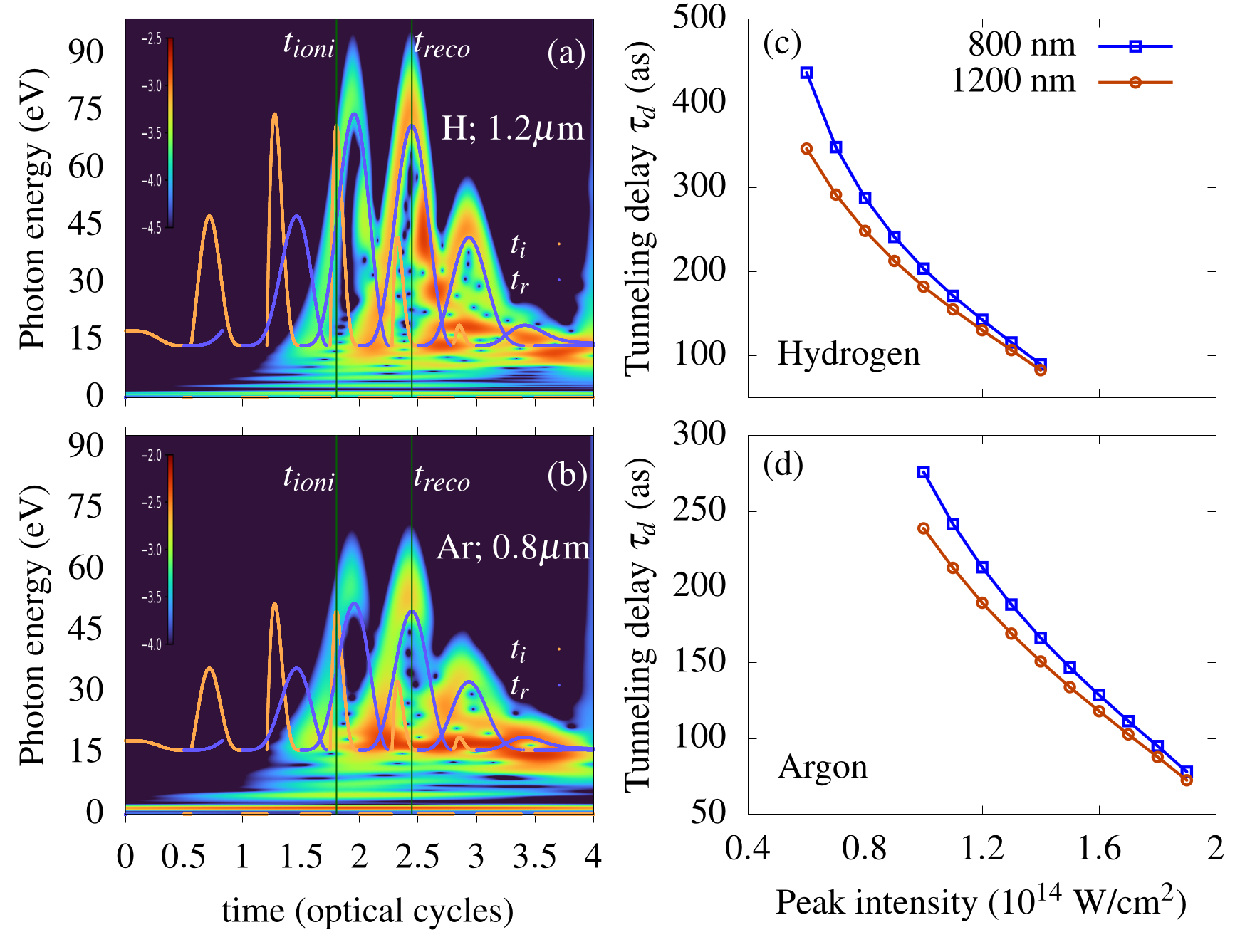}
   \caption{The time-frequency and classical trajectory analysis of the HHG for Hydrogen with 1200 nm (1.5$\times 10^{14}$ W/cm$^2$) (a), and for the Argon with 800 nm (2$\times 10^{14}$ W/cm$^2$) laser (b) are presented. The variation of the tunneling delay with the peak intensity are compared for three different wavelengths for Hydrogen (c) and Argon (d). The solid lines in (c) and (d) represent the $\propto 1/\sqrt{I_0}$ scaling.}
   \label{fig4}
 \end{figure}

\subsection{Comparison with other atomic species}

Figure~\ref{fig4} presents a comparative time--frequency and classical trajectory analysis of high-harmonic generation (HHG) for different atomic species, emphasizing the role of adiabatic tunneling dynamics and its dependence on the laser parameters. Figures~\ref{fig4}(a) and~\ref{fig4}(b) show the Gabor time--frequency maps together with the corresponding classical electron trajectories for Hydrogen driven by a 1200~nm laser at an intensity of $1.5\times10^{14}\,\mathrm{W/cm^2}$ and for Argon driven by an 800~nm laser at $2\times10^{14}\,\mathrm{W/cm^2}$, respectively. In both cases, the emission times inferred from the time--frequency analysis closely follow the classical trajectories, indicating that wavepacket dispersion is not prominent for the chosen laser parameters. This agreement justifies the applicability of the classical three-step model (TSM) for extracting tunneling delays, in line with observations reported by other experimental and theoretical studies.

The comparison across atomic species highlights the pronounced role of the adiabatic approximation for Hydrogen and Argon, whose lower ionization potentials make the effect of finite tunneling time more visible than in Helium. Figures~\ref{fig4}(c) and~\ref{fig4}(d) show the variation of the tunneling delay $\tau_d$ with the peak laser intensity for three different wavelengths, presented separately for Hydrogen and Argon. A systematic $\propto 1/\sqrt{I_0}$ intensity dependence is also observed in both cases, demonstrating the robustness of the extracted delays against changes in wavelength while preserving the expected scaling behavior.

It is worth emphasizing that existing attoclock measurements for Hydrogen largely correspond to the over-the-barrier ionization regime, which cannot be addressed within the present framework. Since the definition of a tunneling delay loses its meaning once the barrier is fully suppressed, the present analysis is deliberately restricted to laser intensities for which tunneling remains the dominant ionization mechanism. Within this regime, Fig.~\ref{fig4} establishes that the classical description, corroborated by the TDSE-based time--frequency analysis, provides a consistent and physically transparent basis for discussing tunneling delays in different atomic species.

\subsection{Effect of the peak field amplitude}
 
\begin{figure}[b]
  \includegraphics[width=\columnwidth]{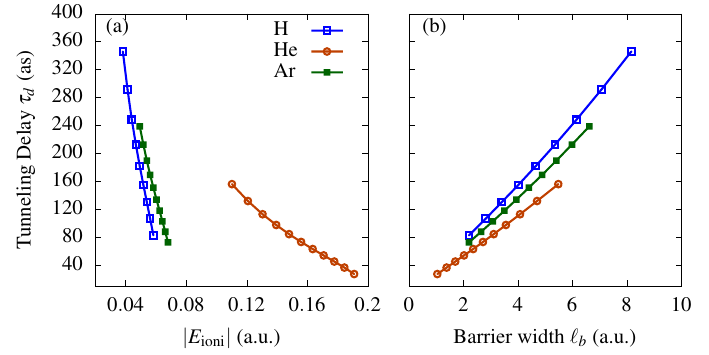}
   \caption{Variation of the tunneling delay with the electric field strength at ionization ($|E_\text{ioni}|$) (a) and the barrier width (b) are compared for Hydrogen, Helium and Argon. For all the cases the wavelength is considered to be 1200 nm, and the peak intensity is varied from $0.5\times 10^{14} - 1.5\times 10^{14}$ W/cm$^2$, and accordingly the $|E_\text{ioni}|$ and so the $\ell_b$ is estimated.} 
   \label{fig5}
 \end{figure}

 Figure~\ref{fig5} examines the dependence of the tunneling delay on the instantaneous field conditions at the moment of ionization, thereby isolating the role of the peak electric field amplitude from purely intensity-based trends. Figure~\ref{fig5}(a) shows the variation of the tunneling delay $\tau_d$ as a function of the electric field strength at ionization, $|E_{\text{ioni}}|$, for Hydrogen, Helium, and Argon. For all species, the wavelength is fixed at 1200~nm while the peak intensity is varied in the range $0.5\times10^{14}$ to $1.5\times10^{14}$~W/cm$^2$. As the peak intensity increases, the instantaneous field experienced by the electron at the ionization time also increases, leading to a systematic reduction in the tunneling delay. This monotonic trend reflects the enhanced adiabaticity of the ionization process at stronger fields, where the electron traverses the barrier more rapidly.

As shown in Fig.~\ref{fig5}(a), the tunneling delay $\tau_d$ decreases monotonically with increasing instantaneous field strength $|E_{\mathrm{ioni}}|$ for Hydrogen, Helium, and Argon. This behavior reflects the fact that a stronger local field suppresses the Coulomb barrier more efficiently, reducing both its width and the time required for the electron to traverse it. Importantly, the use of $|E_{\mathrm{ioni}}|$ accounts for the sub-cycle nature of ionization and removes ambiguities associated with comparing different laser pulses solely on the basis of peak intensity. The observed trend therefore provides direct evidence that the tunneling delay is governed by the instantaneous field configuration at the moment of ionization.

The solid lines in Fig.~\ref{fig5}(a) represent fits proportional to $1/|E_{\mathrm{ioni}}|$ for all three atomic species. This scaling is consistent with the expected strong-field behavior, since the laser intensity scales as $I_0 \propto E_0^2$, implying $1/\sqrt{I_0} \propto 1/E_0$. The observation that the extracted tunneling-delay diagnostic follows a $1/|E_{\mathrm{ioni}}|$ dependence confirms that the dominant field dependence of $\tau_d$ is governed by the instantaneous electric field at the ionization time rather than by the cycle-averaged peak intensity alone. While this scaling is anticipated within semiclassical models of tunneling, its emergence here from a TDSE-guided HHG analysis provides an important internal consistency check and motivates the subsequent interpretation in terms of barrier geometry.

Figure~\ref{fig5}(b) further sharpens this interpretation by recasting the same data in terms of the effective barrier width $\ell_b$, which is determined by the instantaneous field strength and the ionization potential of the target. When plotted as a function of $\ell_b$, the tunneling-delay data for all three atomic species collapse onto a nearly universal curve. This collapse demonstrates that the barrier width is the dominant parameter controlling the tunneling dynamics, while atomic-specific effects enter primarily through their influence on the barrier geometry. In particular, targets with lower ionization potentials, such as Hydrogen and Argon, correspond to narrower barriers for a given field strength and consequently exhibit shorter tunneling delays than Helium.

The significance of Fig.~\ref{fig5} lies in establishing a unified, geometry-based interpretation of the tunneling delay that transcends species-specific details and laser-intensity scaling alone. By explicitly linking $\tau_d$ to the instantaneous barrier width, this analysis reinforces the adiabatic picture of tunneling adopted in the present work and provides a physically transparent framework for comparing different atomic systems on equal footing. Within the parameter regime considered, Fig.~\ref{fig5} thus identifies the local barrier geometry at the ionization time as the primary quantity governing the effective tunneling delay.

 \section{Results Summary}
 \label{sec4}
 
 \begin{figure}[b]
   \includegraphics[width=\columnwidth]{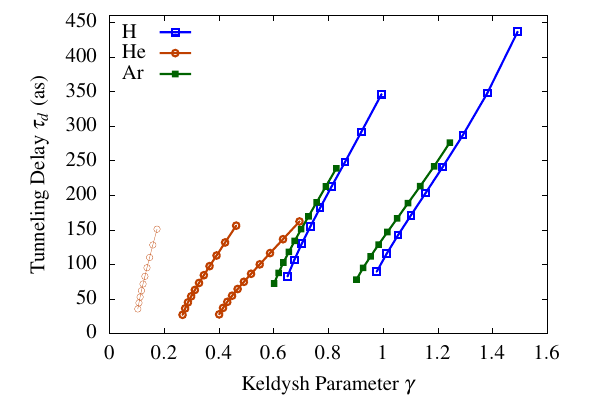}
   \caption{The tunneling delay for Hydrogen, Helium and Argon are complied and presented in terms of the Keldysh parameter $\gamma$, which depends on the peak intensity, and the wavelength of the laser. For Helium atom we have also shown the variation of $\tau_d$ ($\gamma < 0.2$) in a light color signifying the results for 3200 nm wavelength, keeping all other laser parameters are same. Actually for such a long wavelength the wavefunction dispersion is pretty large and the HHG spectra is not clean. However, we have used a sub-cycle pulse of 3200 nm and the predicted tunneling delay (boxed data point) is found to be consistent with this curve. }
   \label{fig6}
 \end{figure}
 
Figure~\ref{fig6} summarizes the main results of this work by compiling the tunneling delay $\tau_d$ for Hydrogen, Helium, and Argon and presenting them as a function of the Keldysh parameter $\gamma$. Since $\gamma$ depends explicitly on both the laser wavelength and the peak intensity, this representation provides a unified framework to compare different atomic species and laser parameters on the same footing. Across all three atoms, the tunneling delay shows a systematic dependence on $\gamma$, decreasing as the system moves deeper into the adiabatic tunneling regime ($\gamma \ll 1$). This collapse of data highlights the central role of the adiabaticity parameter in governing the tunneling dynamics, largely independent of the specific atomic species.

\begin{table*}[t]
\caption{Comparison of effective tunneling-delay estimates obtained in the present HHG-based diagnostic framework with representative delay values inferred from attoclock measurements reported in the literature. Attoclock values are model-dependent and typically extracted from angular offsets using Coulomb-corrected or quantum-trajectory models. The comparison is intended to highlight consistency in magnitude and scaling rather than one-to-one numerical agreement.}
\label{tab:attoclock_comparison}
\begin{ruledtabular}
\begin{tabular}{llcccc}
Target & Reference & $\lambda$ (nm) & $I_0$ ($10^{14}$ W/cm$^2$) & $\gamma$ & Delay (as)  \\ \hline
H & Sainadh \emph{et al.} \cite{Sainadh2019} & 800 -- 1300 & 1.0 -- 3.0  & $\sim$ 0.5 -- 1.0 & $\lesssim$ 50 -- 100   \\
H & this work & 800 -- 1200 & 0.5 -- 1.5 & $\sim$ 0.4 -- 0.9 & 40 -- 90   \\[2pt]

He & Pfeiffer \emph{et al.} \cite{Pfeiffer2012} & 800 & 1.5 -- 3.0 & $\sim$ 0.7 -- 1.2 & 60 -- 120    \\
He & Camus \emph{et al.} \cite{Camus2020} & 800 -- 1300 & 1.0 -- 2.0 & $\sim$ 0.5 -- 1.0 & 50 -- 110   \\
He & this work & 800 -- 1200 & 0.5 -- 1.5 & $\sim$ 0.3 -- 0.8 & 60 -- 130   \\[2pt]

Ar & Pfeiffer \emph{et al.} \cite{Pfeiffer2012} & 800 & 1.0 -- 2.5 & $\sim$ 0.6 -- 1.1 & 80 -- 150   \\
Ar & Ni \emph{et al.} \cite{Ni2020PRL} & 800 -- 1300 & 0.8 -- 2.0 & $\sim$ 0.4 -- 1.0 & 70 -- 140   \\
Ar & this work & 800 -- 1200 & 0.5 -- 2.0 & $\sim$ 0.3 -- 0.9 & 70 -- 160   
\end{tabular}
\end{ruledtabular}
\end{table*}

For Helium, additional data points corresponding to a long wavelength of 3200~nm are shown in a lighter color for $\gamma<0.2$. At such long wavelengths, the electronic wavepacket undergoes significant spatial dispersion, leading to a degradation of the HHG spectral clarity under standard multi-cycle driving conditions. Consequently, a clean time--frequency analysis becomes challenging. To address this limitation, a sub-cycle 3200~nm pulse is employed, and the corresponding tunneling delay is indicated by the light-colored circles. Despite the large wavefunction dispersion inherent to long-wavelength driving, the extracted tunneling delay remains consistent with the overall $\tau_d$--$\gamma$ trend established by the shorter-wavelength data.

Overall, Fig.~\ref{fig6} provides a consolidated view of tunneling delays across different atoms, wavelengths, and intensities, demonstrating that the Keldysh parameter serves as an effective universal descriptor of the tunneling dynamics. This final comparison reinforces the internal consistency of the present analysis and supports the interpretation of tunneling delay as an adiabatic, barrier-controlled process within the tunneling regime.

In Table~\ref{tab:attoclock_comparison}, we compare the effective tunneling-delay estimates obtained in the present HHG-based diagnostic framework with representative delay values inferred from attoclock measurements reported in the literature for Hydrogen, Helium, and Argon. The attoclock values correspond to model-dependent delays extracted from angular offsets using Coulomb-corrected or quantum-trajectory analyses. The comparison is intended to highlight consistency in magnitude, scaling behavior, and Keldysh-parameter dependence, rather than one-to-one numerical agreement between different approaches.

\section{Concluding Remarks}
\label{sec5}

In conclusion, we have demonstrated that high-harmonic generation, when analyzed through a combined TDSE and classical-trajectory framework, can provide meaningful insight into tunneling delay in strong-field ionization. By correlating time--frequency features of HHG with classical ionization and recombination times, we extracted an effective tunneling delay associated with the traversal of the laser-suppressed Coulomb barrier.

The extracted delays exhibit systematic and physically intuitive trends. In particular, the tunneling delay decreases with increasing peak intensity and instantaneous field strength, scales approximately as $1/\sqrt{I_0}$, and shows a near-universal dependence when expressed in terms of the Keldysh parameter. These observations are consistent with established theoretical expectations and attoclock-based results, lending confidence to the internal consistency of the approach. Furthermore, the collapse of data across different atomic species highlights the dominant role of barrier geometry and adiabaticity in governing tunneling dynamics.

It is important to emphasize that HHG does not constitute a direct measurement of tunneling time. The present approach relies on a classical interpretation of the electron motion under the barrier, guided and validated by TDSE simulations. As such, the inferred tunneling delay should be viewed as an effective diagnostic quantity rather than a uniquely defined observable. Nevertheless, the close agreement with known scaling laws and attoclock trends suggests that HHG provides a valuable and independent perspective on tunneling dynamics.

From a broader standpoint, the results presented here position HHG as a complementary tool in the ongoing effort to understand tunneling time in strong-field physics. By offering an alternative route that is experimentally and theoretically well established, HHG-based analysis can serve as a useful cross-check for attoclock measurements and help disentangle the roles of adiabaticity, Coulomb interaction, and laser parameters. Future extensions to more complex targets and nonadiabatic regimes may further clarify the scope and limitations of this supplementary diagnostic.

\section*{Acknowledgments} The authors acknowledge the Department of Science
and Technology (DST) for providing computational resources through the FIST program (Project No. SR/FST/PS-
1/2017/30). Also authors acknowledge BITS - Pilani, Pilani Campus for providing HPC facility.

%\bibliographystyle{apsrev4-2}
%\bibliography{Bibliography}

\begin{thebibliography}{31}%
\makeatletter
\providecommand \@ifxundefined [1]{%
 \@ifx{#1\undefined}
}%
\providecommand \@ifnum [1]{%
 \ifnum #1\expandafter \@firstoftwo
 \else \expandafter \@secondoftwo
 \fi
}%
\providecommand \@ifx [1]{%
 \ifx #1\expandafter \@firstoftwo
 \else \expandafter \@secondoftwo
 \fi
}%
\providecommand \natexlab [1]{#1}%
\providecommand \enquote  [1]{``#1''}%
\providecommand \bibnamefont  [1]{#1}%
\providecommand \bibfnamefont [1]{#1}%
\providecommand \citenamefont [1]{#1}%
\providecommand \href@noop [0]{\@secondoftwo}%
\providecommand \href [0]{\begingroup \@sanitize@url \@href}%
\providecommand \@href[1]{\@@startlink{#1}\@@href}%
\providecommand \@@href[1]{\endgroup#1\@@endlink}%
\providecommand \@sanitize@url [0]{\catcode `\\12\catcode `\$12\catcode
  `\&12\catcode `\#12\catcode `\^12\catcode `\_12\catcode `\%12\relax}%
\providecommand \@@startlink[1]{}%
\providecommand \@@endlink[0]{}%
\providecommand \url  [0]{\begingroup\@sanitize@url \@url }%
\providecommand \@url [1]{\endgroup\@href {#1}{\urlprefix }}%
\providecommand \urlprefix  [0]{URL }%
\providecommand \Eprint [0]{\href }%
\providecommand \doibase [0]{https://doi.org/}%
\providecommand \selectlanguage [0]{\@gobble}%
\providecommand \bibinfo  [0]{\@secondoftwo}%
\providecommand \bibfield  [0]{\@secondoftwo}%
\providecommand \translation [1]{[#1]}%
\providecommand \BibitemOpen [0]{}%
\providecommand \bibitemStop [0]{}%
\providecommand \bibitemNoStop [0]{.\EOS\space}%
\providecommand \EOS [0]{\spacefactor3000\relax}%
\providecommand \BibitemShut  [1]{\csname bibitem#1\endcsname}%
\let\auto@bib@innerbib\@empty
%</preamble>
\bibitem [{\citenamefont {Corkum}(1993)}]{Corkum1993}%
  \BibitemOpen
  \bibfield  {author} {\bibinfo {author} {\bibfnamefont {P.~B.}\ \bibnamefont
  {Corkum}},\ }\href@noop {} {\bibfield  {journal} {\bibinfo  {journal} {Phys.
  Rev. Lett.}\ }\textbf {\bibinfo {volume} {71}},\ \bibinfo {pages} {1994}
  (\bibinfo {year} {1993})}\BibitemShut {NoStop}%
\bibitem [{\citenamefont {Lewenstein}\ \emph {et~al.}(1994)\citenamefont
  {Lewenstein}, \citenamefont {Balcou}, \citenamefont {Ivanov}, \citenamefont
  {L'Huillier},\ and\ \citenamefont {Corkum}}]{Lewenstein1994}%
  \BibitemOpen
  \bibfield  {author} {\bibinfo {author} {\bibfnamefont {M.}~\bibnamefont
  {Lewenstein}}, \bibinfo {author} {\bibfnamefont {P.}~\bibnamefont {Balcou}},
  \bibinfo {author} {\bibfnamefont {M.~Y.}\ \bibnamefont {Ivanov}}, \bibinfo
  {author} {\bibfnamefont {A.}~\bibnamefont {L'Huillier}},\ and\ \bibinfo
  {author} {\bibfnamefont {P.~B.}\ \bibnamefont {Corkum}},\ }\href@noop {}
  {\bibfield  {journal} {\bibinfo  {journal} {Phys. Rev. A}\ }\textbf {\bibinfo
  {volume} {49}},\ \bibinfo {pages} {2117} (\bibinfo {year}
  {1994})}\BibitemShut {NoStop}%
\bibitem [{\citenamefont {Krausz}\ and\ \citenamefont
  {Ivanov}(2009)}]{KrauszIvanov2009}%
  \BibitemOpen
  \bibfield  {author} {\bibinfo {author} {\bibfnamefont {F.}~\bibnamefont
  {Krausz}}\ and\ \bibinfo {author} {\bibfnamefont {M.}~\bibnamefont
  {Ivanov}},\ }\href@noop {} {\bibfield  {journal} {\bibinfo  {journal} {Rev.
  Mod. Phys.}\ }\textbf {\bibinfo {volume} {81}},\ \bibinfo {pages} {163}
  (\bibinfo {year} {2009})}\BibitemShut {NoStop}%
\bibitem [{\citenamefont {Sainadh}\ \emph
  {et~al.}(2019{\natexlab{a}})\citenamefont {Sainadh}, \citenamefont {Xu},
  \citenamefont {Wang}, \citenamefont {Atia-Tul-Noor}, \citenamefont {Wallace},
  \citenamefont {Douguet}, \citenamefont {Bray}, \citenamefont {Ivanov},
  \citenamefont {Bartschat}, \citenamefont {Kheifets}, \citenamefont {Sang},\
  and\ \citenamefont {Litvinyuk}}]{Sainadh2019}%
  \BibitemOpen
  \bibfield  {author} {\bibinfo {author} {\bibfnamefont {U.~S.}\ \bibnamefont
  {Sainadh}}, \bibinfo {author} {\bibfnamefont {H.}~\bibnamefont {Xu}},
  \bibinfo {author} {\bibfnamefont {X.}~\bibnamefont {Wang}}, \bibinfo {author}
  {\bibfnamefont {A.}~\bibnamefont {Atia-Tul-Noor}}, \bibinfo {author}
  {\bibfnamefont {W.~C.}\ \bibnamefont {Wallace}}, \bibinfo {author}
  {\bibfnamefont {N.}~\bibnamefont {Douguet}}, \bibinfo {author} {\bibfnamefont
  {A.}~\bibnamefont {Bray}}, \bibinfo {author} {\bibfnamefont {I.}~\bibnamefont
  {Ivanov}}, \bibinfo {author} {\bibfnamefont {K.}~\bibnamefont {Bartschat}},
  \bibinfo {author} {\bibfnamefont {A.}~\bibnamefont {Kheifets}}, \bibinfo
  {author} {\bibfnamefont {R.~T.}\ \bibnamefont {Sang}},\ and\ \bibinfo
  {author} {\bibfnamefont {I.~V.}\ \bibnamefont {Litvinyuk}},\ }\href
  {https://doi.org/10.1038/s41586-019-1028-3} {\bibfield  {journal} {\bibinfo
  {journal} {Nature}\ }\textbf {\bibinfo {volume} {568}},\ \bibinfo {pages}
  {75} (\bibinfo {year} {2019}{\natexlab{a}})}\BibitemShut {NoStop}%
\bibitem [{\citenamefont {Sainadh}\ \emph
  {et~al.}(2019{\natexlab{b}})\citenamefont {Sainadh}, \citenamefont {Xu},
  \citenamefont {Wang}, \citenamefont {Atia-Tul-Noor}, \citenamefont {Wallace},
  \citenamefont {Douguet}, \citenamefont {Bray}, \citenamefont {Ivanov},
  \citenamefont {Bartschat}, \citenamefont {Kheifets}, \citenamefont {Sang},\
  and\ \citenamefont {Litvinyuk}}]{Sainadh2019_new}%
  \BibitemOpen
  \bibfield  {author} {\bibinfo {author} {\bibfnamefont {U.~S.}\ \bibnamefont
  {Sainadh}}, \bibinfo {author} {\bibfnamefont {H.}~\bibnamefont {Xu}},
  \bibinfo {author} {\bibfnamefont {X.}~\bibnamefont {Wang}}, \bibinfo {author}
  {\bibfnamefont {A.}~\bibnamefont {Atia-Tul-Noor}}, \bibinfo {author}
  {\bibfnamefont {W.~C.}\ \bibnamefont {Wallace}}, \bibinfo {author}
  {\bibfnamefont {N.}~\bibnamefont {Douguet}}, \bibinfo {author} {\bibfnamefont
  {A.}~\bibnamefont {Bray}}, \bibinfo {author} {\bibfnamefont {I.}~\bibnamefont
  {Ivanov}}, \bibinfo {author} {\bibfnamefont {K.}~\bibnamefont {Bartschat}},
  \bibinfo {author} {\bibfnamefont {A.}~\bibnamefont {Kheifets}}, \bibinfo
  {author} {\bibfnamefont {R.~T.}\ \bibnamefont {Sang}},\ and\ \bibinfo
  {author} {\bibfnamefont {I.~V.}\ \bibnamefont {Litvinyuk}},\ }\href
  {https://doi.org/10.1038/s41586-019-1028-3} {\bibfield  {journal} {\bibinfo
  {journal} {Nature}\ }\textbf {\bibinfo {volume} {568}},\ \bibinfo {pages}
  {75} (\bibinfo {year} {2019}{\natexlab{b}})}\BibitemShut {NoStop}%
\bibitem [{\citenamefont {Torlina}\ \emph {et~al.}(2015)\citenamefont
  {Torlina}, \citenamefont {Morales}, \citenamefont {Kaushal}, \citenamefont
  {Ivanov}, \citenamefont {Kheifets}, \citenamefont {Zielinski}, \citenamefont
  {Scrinzi}, \citenamefont {Muller}, \citenamefont {Sukiasyan}, \citenamefont
  {Ivanov},\ and\ \citenamefont {Smirnova}}]{Torlina2015}%
  \BibitemOpen
  \bibfield  {author} {\bibinfo {author} {\bibfnamefont {L.}~\bibnamefont
  {Torlina}}, \bibinfo {author} {\bibfnamefont {F.}~\bibnamefont {Morales}},
  \bibinfo {author} {\bibfnamefont {J.}~\bibnamefont {Kaushal}}, \bibinfo
  {author} {\bibfnamefont {I.}~\bibnamefont {Ivanov}}, \bibinfo {author}
  {\bibfnamefont {A.}~\bibnamefont {Kheifets}}, \bibinfo {author}
  {\bibfnamefont {A.}~\bibnamefont {Zielinski}}, \bibinfo {author}
  {\bibfnamefont {A.}~\bibnamefont {Scrinzi}}, \bibinfo {author} {\bibfnamefont
  {H.~G.}\ \bibnamefont {Muller}}, \bibinfo {author} {\bibfnamefont
  {S.}~\bibnamefont {Sukiasyan}}, \bibinfo {author} {\bibfnamefont
  {M.}~\bibnamefont {Ivanov}},\ and\ \bibinfo {author} {\bibfnamefont
  {O.}~\bibnamefont {Smirnova}},\ }\href {https://doi.org/10.1038/nphys3340}
  {\bibfield  {journal} {\bibinfo  {journal} {Nature Physics}\ }\textbf
  {\bibinfo {volume} {11}},\ \bibinfo {pages} {503} (\bibinfo {year}
  {2015})}\BibitemShut {NoStop}%
\bibitem [{\citenamefont {Landsman}\ \emph {et~al.}(2014)\citenamefont
  {Landsman}, \citenamefont {Weger}, \citenamefont {Maurer}, \citenamefont
  {Boge}, \citenamefont {Ludwig}, \citenamefont {Heuser}, \citenamefont
  {Cirelli}, \citenamefont {Gallmann},\ and\ \citenamefont
  {Keller}}]{Landsman_14}%
  \BibitemOpen
  \bibfield  {author} {\bibinfo {author} {\bibfnamefont {A.~S.}\ \bibnamefont
  {Landsman}}, \bibinfo {author} {\bibfnamefont {M.}~\bibnamefont {Weger}},
  \bibinfo {author} {\bibfnamefont {J.}~\bibnamefont {Maurer}}, \bibinfo
  {author} {\bibfnamefont {R.}~\bibnamefont {Boge}}, \bibinfo {author}
  {\bibfnamefont {A.}~\bibnamefont {Ludwig}}, \bibinfo {author} {\bibfnamefont
  {S.}~\bibnamefont {Heuser}}, \bibinfo {author} {\bibfnamefont
  {C.}~\bibnamefont {Cirelli}}, \bibinfo {author} {\bibfnamefont
  {L.}~\bibnamefont {Gallmann}},\ and\ \bibinfo {author} {\bibfnamefont
  {U.}~\bibnamefont {Keller}},\ }\href
  {https://doi.org/10.1364/OPTICA.1.000343} {\bibfield  {journal} {\bibinfo
  {journal} {Optica}\ }\textbf {\bibinfo {volume} {1}},\ \bibinfo {pages} {343}
  (\bibinfo {year} {2014})}\BibitemShut {NoStop}%
\bibitem [{\citenamefont {Pfeiffer}\ \emph {et~al.}(2012)\citenamefont
  {Pfeiffer}, \citenamefont {Cirelli}, \citenamefont {Smolarski}, \citenamefont
  {Dimitrovski}, \citenamefont {Abu-samha}, \citenamefont {Madsen},\ and\
  \citenamefont {Keller}}]{Pfeiffer2012}%
  \BibitemOpen
  \bibfield  {author} {\bibinfo {author} {\bibfnamefont {A.~N.}\ \bibnamefont
  {Pfeiffer}}, \bibinfo {author} {\bibfnamefont {C.}~\bibnamefont {Cirelli}},
  \bibinfo {author} {\bibfnamefont {M.}~\bibnamefont {Smolarski}}, \bibinfo
  {author} {\bibfnamefont {D.}~\bibnamefont {Dimitrovski}}, \bibinfo {author}
  {\bibfnamefont {M.}~\bibnamefont {Abu-samha}}, \bibinfo {author}
  {\bibfnamefont {L.~B.}\ \bibnamefont {Madsen}},\ and\ \bibinfo {author}
  {\bibfnamefont {U.}~\bibnamefont {Keller}},\ }\href
  {https://doi.org/10.1038/nphys2125} {\bibfield  {journal} {\bibinfo
  {journal} {Nature Physics}\ }\textbf {\bibinfo {volume} {8}},\ \bibinfo
  {pages} {76} (\bibinfo {year} {2012})}\BibitemShut {NoStop}%
\bibitem [{\citenamefont {C.~Hofmann}\ and\ \citenamefont
  {Keller}(2019)}]{Hofmann_2019_JMO}%
  \BibitemOpen
  \bibfield  {author} {\bibinfo {author} {\bibfnamefont {A.~S.~L.}\
  \bibnamefont {C.~Hofmann}}\ and\ \bibinfo {author} {\bibfnamefont
  {U.}~\bibnamefont {Keller}},\ }\href
  {https://doi.org/10.1080/09500340.2019.1596325} {\bibfield  {journal}
  {\bibinfo  {journal} {Journal of Modern Optics}\ }\textbf {\bibinfo {volume}
  {66}},\ \bibinfo {pages} {1052} (\bibinfo {year} {2019})}\BibitemShut
  {NoStop}%
\bibitem [{\citenamefont {Camus}\ and\ \citenamefont
  {et~al.}(2020)}]{Camus2020}%
  \BibitemOpen
  \bibfield  {author} {\bibinfo {author} {\bibfnamefont {N.}~\bibnamefont
  {Camus}}\ and\ \bibinfo {author} {\bibnamefont {et~al.}},\ }\href
  {https://doi.org/10.1103/PhysRevLett.125.123201} {\bibfield  {journal}
  {\bibinfo  {journal} {Phys. Rev. Lett.}\ }\textbf {\bibinfo {volume} {125}},\
  \bibinfo {pages} {123201} (\bibinfo {year} {2020})}\BibitemShut {NoStop}%
\bibitem [{\citenamefont {Ni}\ \emph {et~al.}(2020)\citenamefont {Ni},
  \citenamefont {Saalmann},\ and\ \citenamefont {Rost}}]{Ni2020PRL}%
  \BibitemOpen
  \bibfield  {author} {\bibinfo {author} {\bibfnamefont {H.}~\bibnamefont
  {Ni}}, \bibinfo {author} {\bibfnamefont {U.}~\bibnamefont {Saalmann}},\ and\
  \bibinfo {author} {\bibfnamefont {J.~M.}\ \bibnamefont {Rost}},\ }\href
  {https://doi.org/10.1103/PhysRevLett.125.073202} {\bibfield  {journal}
  {\bibinfo  {journal} {Phys. Rev. Lett.}\ }\textbf {\bibinfo {volume} {125}},\
  \bibinfo {pages} {073202} (\bibinfo {year} {2020})}\BibitemShut {NoStop}%
\bibitem [{\citenamefont {Ni}\ \emph {et~al.}(2021)\citenamefont {Ni},
  \citenamefont {Saalmann},\ and\ \citenamefont {Rost}}]{Ni2021NatCommun}%
  \BibitemOpen
  \bibfield  {author} {\bibinfo {author} {\bibfnamefont {H.}~\bibnamefont
  {Ni}}, \bibinfo {author} {\bibfnamefont {U.}~\bibnamefont {Saalmann}},\ and\
  \bibinfo {author} {\bibfnamefont {J.~M.}\ \bibnamefont {Rost}},\ }\href
  {https://doi.org/10.1038/s41467-021-23885-5} {\bibfield  {journal} {\bibinfo
  {journal} {Nat. Commun.}\ }\textbf {\bibinfo {volume} {12}},\ \bibinfo
  {pages} {3506} (\bibinfo {year} {2021})}\BibitemShut {NoStop}%
\bibitem [{\citenamefont {Ni}\ \emph {et~al.}(2016)\citenamefont {Ni},
  \citenamefont {Saalmann},\ and\ \citenamefont
  {Rost}}]{PhysRevLett.117.023002}%
  \BibitemOpen
  \bibfield  {author} {\bibinfo {author} {\bibfnamefont {H.}~\bibnamefont
  {Ni}}, \bibinfo {author} {\bibfnamefont {U.}~\bibnamefont {Saalmann}},\ and\
  \bibinfo {author} {\bibfnamefont {J.-M.}\ \bibnamefont {Rost}},\ }\href
  {https://doi.org/10.1103/PhysRevLett.117.023002} {\bibfield  {journal}
  {\bibinfo  {journal} {Phys. Rev. Lett.}\ }\textbf {\bibinfo {volume} {117}},\
  \bibinfo {pages} {023002} (\bibinfo {year} {2016})}\BibitemShut {NoStop}%
\bibitem [{\citenamefont {Rost}\ and\ \citenamefont
  {Saalmann}(2019)}]{RostSaalmann2019}%
  \BibitemOpen
  \bibfield  {author} {\bibinfo {author} {\bibfnamefont {J.~M.}\ \bibnamefont
  {Rost}}\ and\ \bibinfo {author} {\bibfnamefont {U.}~\bibnamefont
  {Saalmann}},\ }\href {https://doi.org/10.1088/1361-6455/ab2f33} {\bibfield
  {journal} {\bibinfo  {journal} {J. Phys. B}\ }\textbf {\bibinfo {volume}
  {52}},\ \bibinfo {pages} {163001} (\bibinfo {year} {2019})}\BibitemShut
  {NoStop}%
\bibitem [{\citenamefont {Kaushal}\ \emph {et~al.}(2015)\citenamefont
  {Kaushal}, \citenamefont {Morales},\ and\ \citenamefont
  {Smirnova}}]{Kaushal2016_PRA}%
  \BibitemOpen
  \bibfield  {author} {\bibinfo {author} {\bibfnamefont {J.}~\bibnamefont
  {Kaushal}}, \bibinfo {author} {\bibfnamefont {F.}~\bibnamefont {Morales}},\
  and\ \bibinfo {author} {\bibfnamefont {O.}~\bibnamefont {Smirnova}},\ }\href
  {https://doi.org/10.1103/PhysRevA.92.063405} {\bibfield  {journal} {\bibinfo
  {journal} {Phys. Rev. A}\ }\textbf {\bibinfo {volume} {92}},\ \bibinfo
  {pages} {063405} (\bibinfo {year} {2015})}\BibitemShut {NoStop}%
\bibitem [{\citenamefont {Yakaboylu}\ \emph {et~al.}(2014)\citenamefont
  {Yakaboylu}, \citenamefont {Klaiber},\ and\ \citenamefont
  {Hatsagortsyan}}]{PhysRevA.90.012116}%
  \BibitemOpen
  \bibfield  {author} {\bibinfo {author} {\bibfnamefont {E.}~\bibnamefont
  {Yakaboylu}}, \bibinfo {author} {\bibfnamefont {M.}~\bibnamefont {Klaiber}},\
  and\ \bibinfo {author} {\bibfnamefont {K.~Z.}\ \bibnamefont
  {Hatsagortsyan}},\ }\href {https://doi.org/10.1103/PhysRevA.90.012116}
  {\bibfield  {journal} {\bibinfo  {journal} {Phys. Rev. A}\ }\textbf {\bibinfo
  {volume} {90}},\ \bibinfo {pages} {012116} (\bibinfo {year}
  {2014})}\BibitemShut {NoStop}%
\bibitem [{\citenamefont {Bray}\ \emph {et~al.}(2018)\citenamefont {Bray},
  \citenamefont {Eckart},\ and\ \citenamefont {Kheifets}}]{Bray2018}%
  \BibitemOpen
  \bibfield  {author} {\bibinfo {author} {\bibfnamefont {A.~W.}\ \bibnamefont
  {Bray}}, \bibinfo {author} {\bibfnamefont {S.}~\bibnamefont {Eckart}},\ and\
  \bibinfo {author} {\bibfnamefont {A.~S.}\ \bibnamefont {Kheifets}},\ }\href
  {https://doi.org/10.1103/PhysRevLett.121.123201} {\bibfield  {journal}
  {\bibinfo  {journal} {Phys. Rev. Lett.}\ }\textbf {\bibinfo {volume} {121}},\
  \bibinfo {pages} {123201} (\bibinfo {year} {2018})}\BibitemShut {NoStop}%
\bibitem [{\citenamefont {Pisanty}\ and\ \citenamefont
  {Ivanov}(2020)}]{Pisanty2020}%
  \BibitemOpen
  \bibfield  {author} {\bibinfo {author} {\bibfnamefont {E.}~\bibnamefont
  {Pisanty}}\ and\ \bibinfo {author} {\bibfnamefont {M.}~\bibnamefont
  {Ivanov}},\ }\href {https://doi.org/10.1103/PhysRevA.102.043108} {\bibfield
  {journal} {\bibinfo  {journal} {Phys. Rev. A}\ }\textbf {\bibinfo {volume}
  {102}},\ \bibinfo {pages} {043108} (\bibinfo {year} {2020})}\BibitemShut
  {NoStop}%
\bibitem [{\citenamefont {Eckart}\ \emph {et~al.}(2021)\citenamefont {Eckart},
  \citenamefont {Bray},\ and\ \citenamefont {Kheifets}}]{Eckart2021}%
  \BibitemOpen
  \bibfield  {author} {\bibinfo {author} {\bibfnamefont {S.}~\bibnamefont
  {Eckart}}, \bibinfo {author} {\bibfnamefont {A.~W.}\ \bibnamefont {Bray}},\
  and\ \bibinfo {author} {\bibfnamefont {A.~S.}\ \bibnamefont {Kheifets}},\
  }\href {https://doi.org/10.1103/PhysRevA.103.013113} {\bibfield  {journal}
  {\bibinfo  {journal} {Phys. Rev. A}\ }\textbf {\bibinfo {volume} {103}},\
  \bibinfo {pages} {013113} (\bibinfo {year} {2021})}\BibitemShut {NoStop}%
\bibitem [{\citenamefont {Liu}\ and\ \citenamefont {et~al.}(2021)}]{Liu2021}%
  \BibitemOpen
  \bibfield  {author} {\bibinfo {author} {\bibfnamefont {J.}~\bibnamefont
  {Liu}}\ and\ \bibinfo {author} {\bibnamefont {et~al.}},\ }\href
  {https://doi.org/10.1103/PhysRevA.104.023109} {\bibfield  {journal} {\bibinfo
   {journal} {Phys. Rev. A}\ }\textbf {\bibinfo {volume} {104}},\ \bibinfo
  {pages} {023109} (\bibinfo {year} {2021})}\BibitemShut {NoStop}%
\bibitem [{\citenamefont {Zhang}\ and\ \citenamefont
  {et~al.}(2022)}]{Zhang2022}%
  \BibitemOpen
  \bibfield  {author} {\bibinfo {author} {\bibfnamefont {Q.}~\bibnamefont
  {Zhang}}\ and\ \bibinfo {author} {\bibnamefont {et~al.}},\ }\href
  {https://doi.org/10.1103/PhysRevA.106.023117} {\bibfield  {journal} {\bibinfo
   {journal} {Phys. Rev. A}\ }\textbf {\bibinfo {volume} {106}},\ \bibinfo
  {pages} {023117} (\bibinfo {year} {2022})}\BibitemShut {NoStop}%
\bibitem [{\citenamefont {Smirnova}\ \emph {et~al.}(2009)\citenamefont
  {Smirnova}, \citenamefont {Mairesse}, \citenamefont {Patchkovskii},
  \citenamefont {Dudovich}, \citenamefont {Villeneuve}, \citenamefont
  {Corkum},\ and\ \citenamefont {Ivanov}}]{Smirnova2009}%
  \BibitemOpen
  \bibfield  {author} {\bibinfo {author} {\bibfnamefont {O.}~\bibnamefont
  {Smirnova}}, \bibinfo {author} {\bibfnamefont {Y.}~\bibnamefont {Mairesse}},
  \bibinfo {author} {\bibfnamefont {S.}~\bibnamefont {Patchkovskii}}, \bibinfo
  {author} {\bibfnamefont {N.}~\bibnamefont {Dudovich}}, \bibinfo {author}
  {\bibfnamefont {D.~M.}\ \bibnamefont {Villeneuve}}, \bibinfo {author}
  {\bibfnamefont {P.~B.}\ \bibnamefont {Corkum}},\ and\ \bibinfo {author}
  {\bibfnamefont {M.~Y.}\ \bibnamefont {Ivanov}},\ }\href@noop {} {\bibfield
  {journal} {\bibinfo  {journal} {Nature}\ }\textbf {\bibinfo {volume} {460}},\
  \bibinfo {pages} {972} (\bibinfo {year} {2009})}\BibitemShut {NoStop}%
\bibitem [{\citenamefont {Chiril\u{a}}\ and\ \citenamefont
  {Lein}(2010)}]{Chirala2010}%
  \BibitemOpen
  \bibfield  {author} {\bibinfo {author} {\bibfnamefont {C.~C.}\ \bibnamefont
  {Chiril\u{a}}}\ and\ \bibinfo {author} {\bibfnamefont {M.}~\bibnamefont
  {Lein}},\ }\href@noop {} {\bibfield  {journal} {\bibinfo  {journal} {Phys.
  Rev. A}\ }\textbf {\bibinfo {volume} {82}},\ \bibinfo {pages} {023410}
  (\bibinfo {year} {2010})}\BibitemShut {NoStop}%
\bibitem [{\citenamefont {Tong}\ and\ \citenamefont {Chu}(1997)}]{TONG1997119}%
  \BibitemOpen
  \bibfield  {author} {\bibinfo {author} {\bibfnamefont {X.-M.}\ \bibnamefont
  {Tong}}\ and\ \bibinfo {author} {\bibfnamefont {S.-I.}\ \bibnamefont {Chu}},\
  }\href {https://doi.org/https://doi.org/10.1016/S0301-0104(97)00063-3}
  {\bibfield  {journal} {\bibinfo  {journal} {Chem. Phys.}\ }\textbf {\bibinfo
  {volume} {217}},\ \bibinfo {pages} {119} (\bibinfo {year}
  {1997})}\BibitemShut {NoStop}%
\bibitem [{\citenamefont {Tong}\ and\ \citenamefont {Lin}(2005)}]{Tong_2005}%
  \BibitemOpen
  \bibfield  {author} {\bibinfo {author} {\bibfnamefont {X.~M.}\ \bibnamefont
  {Tong}}\ and\ \bibinfo {author} {\bibfnamefont {C.~D.}\ \bibnamefont {Lin}},\
  }\href {https://doi.org/10.1088/0953-4075/38/15/001} {\bibfield  {journal}
  {\bibinfo  {journal} {J. Phys. B: At. Mol. Opt. Phys.}\ }\textbf {\bibinfo
  {volume} {38}},\ \bibinfo {pages} {2593} (\bibinfo {year}
  {2005})}\BibitemShut {NoStop}%
\bibitem [{\citenamefont {Holkundkar}\ \emph {et~al.}(2023)\citenamefont
  {Holkundkar}, \citenamefont {Rajpoot},\ and\ \citenamefont
  {Bandyopadhyay}}]{Holkundkar2023_PhysLettA}%
  \BibitemOpen
  \bibfield  {author} {\bibinfo {author} {\bibfnamefont {A.~R.}\ \bibnamefont
  {Holkundkar}}, \bibinfo {author} {\bibfnamefont {R.}~\bibnamefont
  {Rajpoot}},\ and\ \bibinfo {author} {\bibfnamefont {J.~N.}\ \bibnamefont
  {Bandyopadhyay}},\ }\href
  {https://doi.org/https://doi.org/10.1016/j.physleta.2023.128645} {\bibfield
  {journal} {\bibinfo  {journal} {Phys. Lett. A}\ }\textbf {\bibinfo {volume}
  {461}},\ \bibinfo {pages} {128645} (\bibinfo {year} {2023})}\BibitemShut
  {NoStop}%
\bibitem [{\citenamefont {Rajpoot}\ and\ \citenamefont
  {Holkundkar}(2023)}]{Rajpoot2023_JPhysB}%
  \BibitemOpen
  \bibfield  {author} {\bibinfo {author} {\bibfnamefont {R.}~\bibnamefont
  {Rajpoot}}\ and\ \bibinfo {author} {\bibfnamefont {A.~R.}\ \bibnamefont
  {Holkundkar}},\ }\href {https://doi.org/10.1088/1361-6455/acc4fb} {\bibfield
  {journal} {\bibinfo  {journal} {J. Phys. B: At. Mol. Opt. Phys.}\ }\textbf
  {\bibinfo {volume} {56}},\ \bibinfo {pages} {105402} (\bibinfo {year}
  {2023})}\BibitemShut {NoStop}%
\bibitem [{\citenamefont {Bransden}\ and\ \citenamefont
  {Joachain}(2003)}]{BransdesnBook}%
  \BibitemOpen
  \bibfield  {author} {\bibinfo {author} {\bibfnamefont {B.~H.}\ \bibnamefont
  {Bransden}}\ and\ \bibinfo {author} {\bibfnamefont {C.~J.}\ \bibnamefont
  {Joachain}},\ }\href@noop {} {\emph {\bibinfo {title} {Physics of atoms and
  molecules}}},\ \bibinfo {edition} {second edition.}\ ed.\ (\bibinfo
  {publisher} {Prentice Hall},\ \bibinfo {address} {Harlow},\ \bibinfo {year}
  {2003})\BibitemShut {NoStop}%
\bibitem [{\citenamefont {Kramida}\ \emph {et~al.}(2023)\citenamefont
  {Kramida}, \citenamefont {{Yu.~Ralchenko}}, \citenamefont {Reader},\ and\
  \citenamefont {{NIST ASD Team}}}]{NIST_ASD}%
  \BibitemOpen
  \bibfield  {author} {\bibinfo {author} {\bibfnamefont {A.}~\bibnamefont
  {Kramida}}, \bibinfo {author} {\bibnamefont {{Yu.~Ralchenko}}}, \bibinfo
  {author} {\bibfnamefont {J.}~\bibnamefont {Reader}},\ and\ \bibinfo {author}
  {\bibnamefont {{NIST ASD Team}}},\ }\href
  {https://doi.org/https://doi.org/10.18434/T4W30F} {\bibfield  {journal}
  {\bibinfo  {journal} {NIST Atomic Spectra Database}\ }\textbf {\bibinfo
  {volume} {ver 5.11}},\ \bibinfo {pages} {online} (\bibinfo {year}
  {2023})}\BibitemShut {NoStop}%
\bibitem [{\citenamefont {van~de Sand}\ and\ \citenamefont
  {Rost}(1999)}]{sandPRL_1999}%
  \BibitemOpen
  \bibfield  {author} {\bibinfo {author} {\bibfnamefont {G.}~\bibnamefont
  {van~de Sand}}\ and\ \bibinfo {author} {\bibfnamefont {J.~M.}\ \bibnamefont
  {Rost}},\ }\href {https://doi.org/10.1103/PhysRevLett.83.524} {\bibfield
  {journal} {\bibinfo  {journal} {Phys. Rev. Lett.}\ }\textbf {\bibinfo
  {volume} {83}},\ \bibinfo {pages} {524} (\bibinfo {year} {1999})}\BibitemShut
  {NoStop}%
\bibitem [{\citenamefont {Arts}\ and\ \citenamefont {van~den
  Broek}(2022)}]{fcwt}%
  \BibitemOpen
  \bibfield  {author} {\bibinfo {author} {\bibfnamefont {L.~P.~A.}\
  \bibnamefont {Arts}}\ and\ \bibinfo {author} {\bibfnamefont {E.~L.}\
  \bibnamefont {van~den Broek}},\ }\href
  {https://doi.org/10.1038/s43588-021-00183-z} {\bibfield  {journal} {\bibinfo
  {journal} {Nat. Comput. Sci.}\ }\textbf {\bibinfo {volume} {2}},\ \bibinfo
  {pages} {47} (\bibinfo {year} {2022})}\BibitemShut {NoStop}%
\end{thebibliography}


\end{document}